\documentclass[9pt,twocolumn,twoside]{osajnl_custom}
\usepackage{graphicx} 
\usepackage{siunitx}
\usepackage{multirow}
\usepackage{balance}
\journal{custom} 

\setboolean{shortarticle}{true}

\title{Measuring Laser Beams with a Neural Network}

\author{Lucas R. Hofer}
\author{Milan Krstaji\'{c}}
\author[1*]{Robert P. Smith}

\affil{Clarendon Laboratory, University of Oxford, Parks Road, Oxford OX1 3PU, United Kingdom}
\affil[*]{Corresponding author: robert.smith@physics.ox.ac.uk}

 \doi{\url{}}

\begin{abstract}
A deep neural network (NN) is used to simultaneously detect laser beams in images and measure their center coordinates, radii and angular orientations. A dataset of images containing simulated laser beams and a dataset of images with experimental laser beams---generated using a spatial light modulator---are used to train and evaluate the NN. After training on the simulated dataset the NN achieves beam parameter root-mean-square-errors (RMSEs) of less than 3.4\% on the experimental dataset. Subsequent training on the experimental dataset causes the RMSEs to fall below 1.1\%. The NN method can be used as a stand-alone measurement of the beam parameters or can compliment other beam profiling methods by providing an accurate region-of-interest.
\end{abstract}

\setboolean{displaycopyright}{false}

\begin{document}
\maketitle
\vspace{.1cm}

\section{Introduction}
\vspace{.05cm}

Profiling multiple laser beams on a single image sensor has become increasingly important due to the growing number of multi-beam applications. Spatial light modulators \cite{konforti1988phase}, for example, can create multiple, dynamically controlled laser beams---used for optical tweezer arrays in cold atom experiments \cite{Barredo, Mello, Endres} and multi-site neuron activation in two-photon microscopy \cite{Nikolenko}---while diffractive optical elements allow multiple beams to be created for machining applications \cite{ma13132962, katz2018using} and can also form laser beam arrays used in medical skin treatment procedures \cite{tanghetti2016histology, lee2019pattern}.  

In recent years deep neural networks (NNs) have been applied with great success to the analysis of scientific image data. Convolutional neural networks CNNs \cite{krizhevsky2012imagenet, resnet} often form the basis for image analysis NNs and have been used within optics for tasks such as laser beam mode classification \cite{doster2017machine, hofer1}, modal decomposition \cite{lohani2018use, an2020fast, Schiworski21} and determination of a beam's center coordinates \cite{lin2018application}. Object detection neural networks (ODNN) \cite{fasterrcnn, redmon2018yolov3}, which are based on CNNs, can detect objects in images, classify the objects \cite{hofer2021atom} and determine regions-of-interest (ROIs) which bound the objects. In this work, an ODNN \cite{rrpnn} which returns rotated regions-of-interest (RROIs) is used to identify multiple TEM$_{00}$ Gaussian laser beams in images and simultaneously measure all their spatial parameters.

\begin{figure}[t!]
\centering
  \includegraphics[scale=0.85]{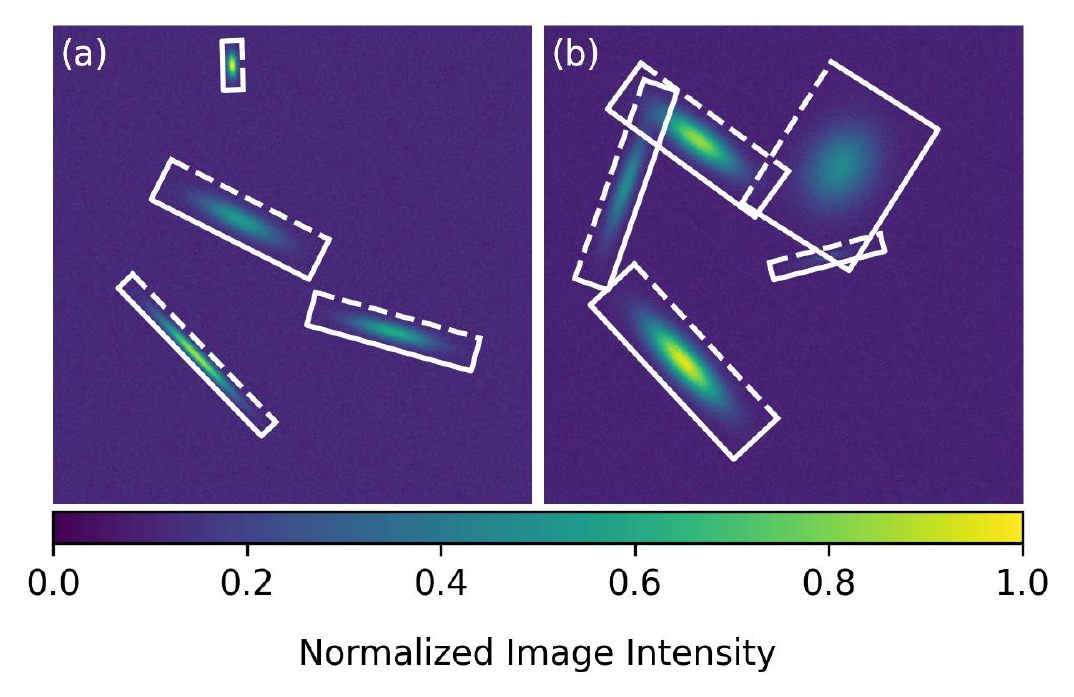}
  \caption{\label{odnn:fig:sim_data} Simulated images. (a)-(b) Images sampled from the simulated dataset with a rotated region-of-interest (RROI) box plotted around each laser beam. The tops of the RROI boxes lie parallel to the major axis and are denoted with dashed lines.}
\end{figure}

The intensity distribution $I(x, y)$ for a TEM$_{00}$ Gaussian beam (see Fig.~\ref{odnn:fig:sim_data}) in a plane orthogonal to the beam's axis of propagation is given by

\begin{equation}
I(x, y) = I_0 e^{-2\left[\frac{\left[\left(x-x_0\right)\cos\theta+\left(y-y_0\right)\sin\theta\right]^2}{w_x^2}+\frac{\left[\left(y-y_0\right)\cos\theta-\left(x-x_0\right)\sin\theta\right]^2}{w_y^2}\right]}\text{,}
\label{odnn:eq:2dgauss}
\end{equation}

\noindent where $I_0$ is the peak intensity of the beam, $x_0$ and $y_0$ are the beam's center coordinates, $w_x$ and $w_y$ are the major and minor radii and $\theta$ is the angular orientation. Although higher-order modes (e.g. Hermite-Gaussian) have experimental applications \cite{Gaunt2013}, we exclusively focus on the TEM$_{00}$ Gaussian beams---henceforth referred to simply as Gaussian beams. The majority of laser beams used in both research and industrial applications have a mode content composed primarily of the TEM$_{00}$ mode thus making it a good approximation for a beam's intensity distribution.

There are several standarized ways to measure Gaussian beams including scanning slit \cite{mccally1984measurement}, knife edge \cite{siegman1991choice} and camera based methods \cite{siegman1998maybe}. Within the camera based methods, the second moment measurement \cite{ross2013laser} is the industry standard \cite{iso2005lasers} as it allows for fast calculation of multi-modal beams. However, the second moment method is prone to statistical error from image noise \cite{Hofer2017} and several standardized methods are used to combat this including thresholding low-intensity pixels and performing calculations within a ROI centered on the beam \cite{iso2005lasers}.

When profiling Gaussian beams, a two-dimensional (2D) fit of the Gaussian beam to Eq.~\ref{odnn:eq:2dgauss} can also be used due to \textit{a priori} knowledge of the beam mode. A properly chosen ROI  increases the 2D fit accuracy by removing portions of the image without relevant data and also decreases the calculation time by fitting a smaller area. Even with an appropriate ROI, the 2D fit is significantly slower than the second moment, but allows for higher accuracy---particularly in noisy images.

For both the second moment and 2D fit methods the ROI is generally found via an iterative method such as calculating the beam width within a ROI, recalculating the ROI using this value and then repeating the process until the beam width converges \cite{iso2005lasers}. However, iterative methods are computationally expensive, have difficulty converging if the image noise is too high and are generally applicable when only a single laser beam is present in the image.

We present a deep neural network based method that allows for an arbitrary number of beams on a single image to be detected and their spatial parameters \{$x_0$, $y_0$, $w_x$, $w_y$, $\theta$\} determined simultaneously, which significantly simplifies the laser beam analysis pipeline. If either the second moment or 2D fit of the beam is still required (e.g. for an ISO 11146 \cite{iso2005lasers} compliant beam measurement), the spatial parameters returned by the NN can be used to determine, ROIs, RROIs or elliptical RROIs in which these calculations can be performed. Furthermore, this method allows for measurement of Gaussian beams with overlapping edges---which cannot be done with the second moment method and would require prior knowledge of the number of beams for the 2D fit method. 

This paper is organized as follows: section~\ref{odnn:sec:RRPN} describes the NN model used to detect the laser beams and measure their spatial parameters, section~\ref{odnn:sec:simulated} and section~\ref{odnn:sec:experimental} explain how the simulated and experimental datasets are created. Finally, section~\ref{odnn:sec:training_eval} discusses training the NN and the accuracy achieved for both detection and determination of the beams' spatial parameters.

\section{Rotated region proposal neural network}\label{odnn:sec:RRPN}

Although the Gaussian equation includes an intensity parameter (amplitude $I_0$), beam profiling is generally only concerned with the shape and location of the laser beam which can be described by the spatial parameters \{$x_0$, $y_0$, $w_x$, $w_y$, $\theta$\}. The goal is therefore to first detect each laser beam (object) in the image and then measure (regress) their geometric parameters. 

Object detection neural networks have been heavily researched in the last decade with several different architectures developed. Region-CNN (RCNN) \cite{girshick2014rich} class NNs are extremely popular and utilize a convolutional neural network base (CNN) followed by a region-proposal-network (RPN) which returns rough ROIs where objects are likely located. The CNN's output is cropped and pooled using a ROI pooling/alignment stage and passed into one or more classification/regression branches---one of which regresses the ROI coordinates to yield a more accurate value. Although this could be useful for detecting beams, the ROI is aligned along the image axes and only yields information about the center coordinates \{$x_0$, $y_0$\} and the projection of the beam radii on the image axes.

\begin{figure}[h!]
\centering
  \includegraphics[scale=0.125]{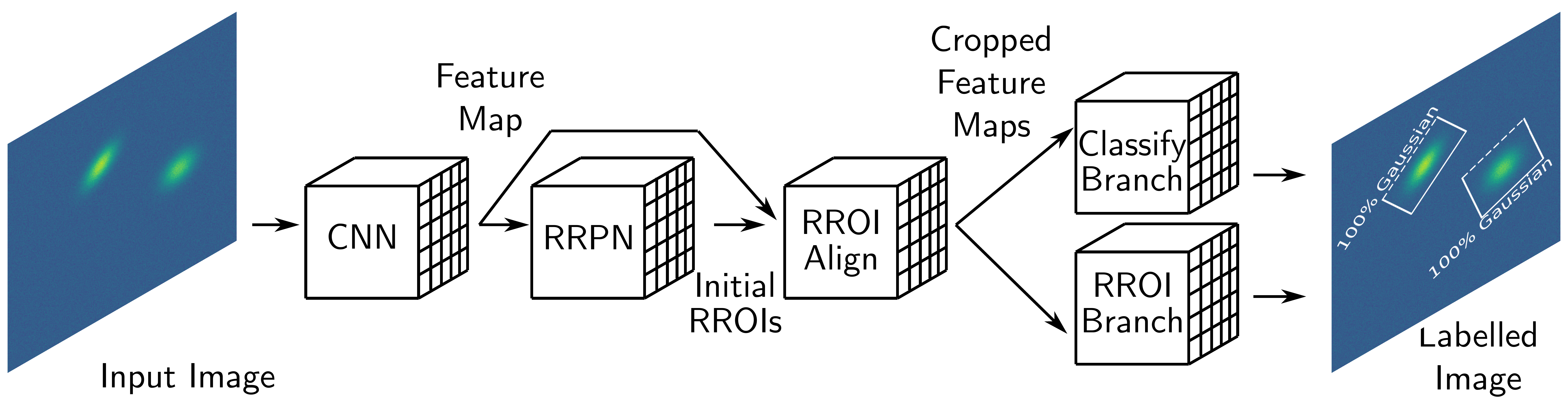}
  \caption{\label{odnn:fig:beam_nn}Rotated Region Proposal Networks model. The neural network begins with a convolutional neural network (CNN) base which returns a feature map of its input. The feature map is passed into the rotated region proposal network (RRPN) which gives a set of rough rotated regions-of-interest (RROIs) where beams are likely located. These RROIs are used to crop and align the CNN's feature map to fixed dimensions in the RROI alignment stage after which the fixed-size feature maps are passed into two parallel branches. The first branch classifies the object and assigns a score to its prediction, whereas the second returns a more accurate RROI.}
\end{figure}

To regress all the beam's geometric parameters we use the Rotated Region Proposal Network (RRPN) \cite{rrpnn} which was initially developed for detecting rotated text in images, but is well suited for detecting laser beams. RRPN is similar to other RCNNs \cite{fasterrcnn}, but returns rotated regions-of-interest (RROIs) rather than ROIs.  RROIs are rotated rectangles which are defined via the center coordinates \{$x_{0r}$, $y_{0r}$\}, widths \{$d_{xr}$, $d_{yr}$\} and angular orientation $\theta_r$\ of the rectangle. For a RROI centered and aligned on a laser beam in an image, the RROI parameters \{$x_{0r}$, $y_{0r}$, $d_{xr}$, $d_{yr}$, $\theta_r$\} directly correspond to the laser beam’s geometric parameters and can thus be rewritten \{$x_0$, $y_0$, $\alpha w_x$, $\alpha w_y$, $\theta$\} where $\alpha$ is a scale factor relating the RROI widths and beam radii. Thus, RRPN can be used to simultaneously detect and measure laser beams.

RRPN (see Fig.~\ref{odnn:fig:beam_nn}) begins with a CNN base (we use ResNet50 \cite{resnet}) which outputs a feature map \cite{zeiler2014visualizing} for a given input. The feature map is then fed into the rotated region proposal network (single node within RRPN) which is similar to the RPN in Faster-RCNN \cite{fasterrcnn}, but returns rough RROIs---rather than ROIs---where objects are likely to be located. The RROIs and feature map are both passed into the RROI alignment stage \cite{huang2018improving} which returns fixed size feature maps via bi-linear interpolation as successive layers require a fixed input. The fixed-size feature maps are then fed into two separate branches: the first classifies the object within the RROI and assigns a score to its prediction, whereas the second does further regression of the RROI parameters.

To train RRPN two datasets are created; the first dataset is comprised of images with simulated Gaussian beams, whereas the second is composed of experimental images with the beams generated using a spatial light modulator. For both the simulated and experimental datasets, the images contain between one and five laser beams, although RRPN could easily be trained to detect a larger number of beams on a single image.

\section{Simulated Dataset}\label{odnn:sec:simulated}

CNN's require diverse image training data to allow them to generalize to new data during inference. However, supervised-learning dataset sizes are often limited due to practical considerations such as the time it takes to manually annotate images. Simulated data allows the annotation bottleneck to be circumvented \cite{wood2016learning, yoo2016pixel, zhang2015learning} as the annotations are calculated directly from the simulation parameters. Since Gaussian beams are relatively easy to simulate, we can create an arbitrarily large dataset filled with unique images by randomizing each beam's Gaussian parameters (see Fig.~\ref{odnn:fig:sim_data}).

When randomizing a beam's parameters, an initial beam radius is first drawn from a uniform distribution with a minimum bound of 5 pixels and a maximum value of 1/6 the 512 pixel image width. An ellipticity value is then drawn from a normal distribution and multiplied by the initial radius to create the second radius value. The larger of the two radii is the major radius $w_x$, the smaller is the minor radius $w_y$ and the angular orientation $\theta$ defines the angle between $w_x$ and the $x$ axis. The angular orientations are randomly chosen between -$\frac{\pi}{2}$ and $\frac{\pi}{2}$---which gives a strict definition the NN can learn to regress while still covering the full range of possible orientations.

The beam's center coordinates are randomly drawn from a uniform distribution, but subject to the constraint that the beam's entire RROI must lie on the simulated sensor surface. Additionally, when more than one beam is present the overlap between beams is restricted to the edges of the distributions. Beam amplitudes are randomly chosen between 0.1 and 1 for all beams and simulated Gaussian noise---with a standard deviation $\sigma_n$ randomly chosen---is added to the image. The background intensity is set to $I_{\text{b}}=2.5\sigma_n$ to prevent the Gaussian noise from being substantially clipped.

Using the process above, a simulated dataset with 5000 simulated images (512$\times$512 pixels, see Fig.~\ref{odnn:fig:sim_data}) is generated---1000 for each beam class (number of beams on the image). The initially monochrome images are normalized and mapped to RGB using the Viridis colormap as the pre-trained NN we use (see Section~\ref{odnn:sec:training_eval}) expects RGB input. The ground truth RROI annotations are calculated from the simulated beam parameters and defined as \{$x_0$, $y_0$, $ \alpha w_x$, $\alpha w_y$, $\theta$\} where we choose $\alpha=3$. Finally, the dataset is randomly split into a training set with 4000 images and validation dataset with 1000 images. 

\section{Experimental dataset}\label{odnn:sec:experimental}

\begin{figure}[t!]
\centering
  \includegraphics[scale=.7]{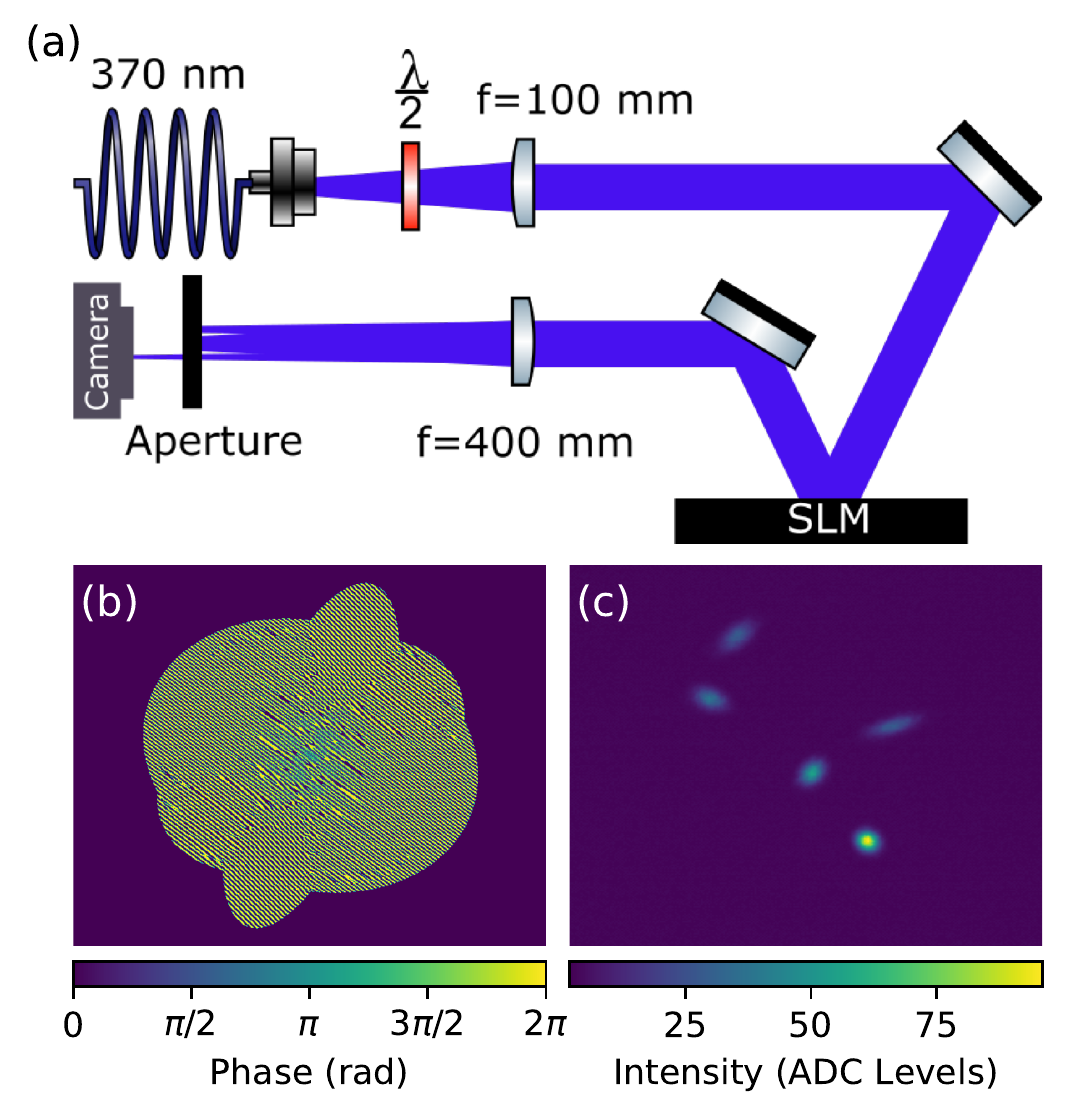}
  \caption{\label{odnn:fig:beam_experiment}Experimental setup. (a) A 370 nm laser beam exits a single-mode fiber and is collimated with a 100 mm lens before being reflected from a Hamamatsu X13267-05 spatial light modulator (SLM)---which features an 800×600 grid with a pixel pitch of 12.5 $\mu$m. A $\lambda/2$ wave plate between the fiber coupler and collimating lens sets the beam's polarization parallel to the SLM's vertical axis as the SLM is polarization sensitive. The reflected beam is focused using a $f$=400 mm lens placed a distance $f$ from the SLM and passes through an aperture placed directly (\textasciitilde1 cm) before the lens' focus---where the different diffraction orders can be resolved. The aperture allows through only the first-order diffracted light which is then imaged by a camera at the Fourier plane. (b) A complex amplitude modulation hologram used to generate Gaussian beams in the Fourier plane for the experimental dataset. (c) Image generated using the hologram in (b).}
\end{figure}

The experimental dataset is created using a spatial light modulator (SLM) \cite{konforti1988phase} which allows structured light \cite{forbes2019structured, Forbes16} to be created using holograms (see Fig.~\ref{odnn:fig:beam_experiment}a). Our setup begins with a MSquared frequency doubled Ti-Sapphire laser which produces 370 nm light and is coupled into a single-mode optical fiber. The beam exits the fiber and is collimated and reflected off the SLM’s surface—after which the beam passes through a $f$=400 mm lens placed a distance $f$ away from the SLM. An aperture is used to select the first-order diffracted light which is subsequently imaged by a camera placed at the lens’ focus (Fourier plane).

Complex amplitude modulation holograms (CAM) \cite{arrizon2007pixelated} allow both the amplitude and phase of the electric field to be modulated using a phase-only SLM (see Fig.~\ref{odnn:fig:beam_experiment}b). The amplitude and phase of the laser beam are encoded into a single hologram \cite{rosales2017shape} along with a blazed grating so that the beam parameters \{$I_0$, $x_0$, $y_0$, $w_x$, $w_y$, $\theta$\} can be dynamically set in the Fourier plane. Adding multiple CAM holograms together---each with different blazed grating frequencies---creates multiple beams in the Fourier plane (see Fig.~\ref{odnn:fig:beam_experiment}c) which can have different sizes, orientations and positions. 

The experimental images contain between one and five beams with the CAM parameters for each beam randomly drawn from a uniform distribution---within physically realizable bounds. The ground truth beam parameters \{$x_0$, $y_0$, $w_x$, $w_y$, $\theta$\} at the Fourier plane were found for each beam by performing a 2D fit within a ROI 2$\times$ the 1/$e^2$ radii; however, for overlapping beams a multi-Gaussian fit was performed. After fitting all the beams in the experimental dataset, the fits were manually inspected\footnote{We also take a histogram of the reduced $\chi^2$ values for all the fitted beams which has a mean of 1.1 and standard deviation of 0.2. This indicates the experimental beams’ intensity distributions are well approximated by the 2D Gaussian.} and subsequently used to calculate the RROIs as \{$x_0$, $y_0$, $ \alpha w_x$, $\alpha w_y$, $\theta$\} where again $\alpha=3$.

The 1050 images in the experimental dataset---210 images for each number of beams---are split into a training dataset with 800 images and a validation dataset with 250 images.  Although the camera has a sensor size of 1280$\times$1024 pixels, the beams in the Fourier plane are incident on a small area of the sensor and the images are cropped to 256$\times$256 pixels. As with the simulated dataset, the image intensities are normalized and mapped to RGB.

\section{Training and evaluation}\label{odnn:sec:training_eval}

Rather than building the NN model from scratch, Facebook Artificial Intelligence Research's (FAIR) Detectron2 \cite{wu2019detectron2} framework is utilized which implements common machine vision models and is written for speedy training and inference. Since Detectron2 only has pre-trained weights for the CNN base, transfer learning \cite{yosinski2014transferable} can only be partially implemented and significantly more images are needed to train the NN. Our strategy is therefore to first pre-train RRPN on the larger, more diverse simulated dataset before (optionally) doing final training on the experimental dataset.

RRPN is trained on the simulation dataset for 120 epochs using a stochastic gradient descent optimizer with an initial learning rate which is decayed four times. The initial learning rate, learning rate decay scalar and the epochs at which the learning rate is decayed are all used as hyperparameters---along with the momentum and batch size. Nominally either random search \cite{bergstra2012random} or Bayesian optimization would be used to tune the hyperparameters; however, due to the large size of the simulated dataset and limited computational resources (we train our NN in a Google Colab \cite{bisong2019google} notebook) we manually set the hyperparameters to sensible values: batch size of 4, learning rate of 0.01, momentum of 0.9, learning rate decay of 0.1 and learning rate decay at epochs \{80, 100, 110, 115\}.

\begin{figure}[t!]
\centering
  \includegraphics[scale=0.6]{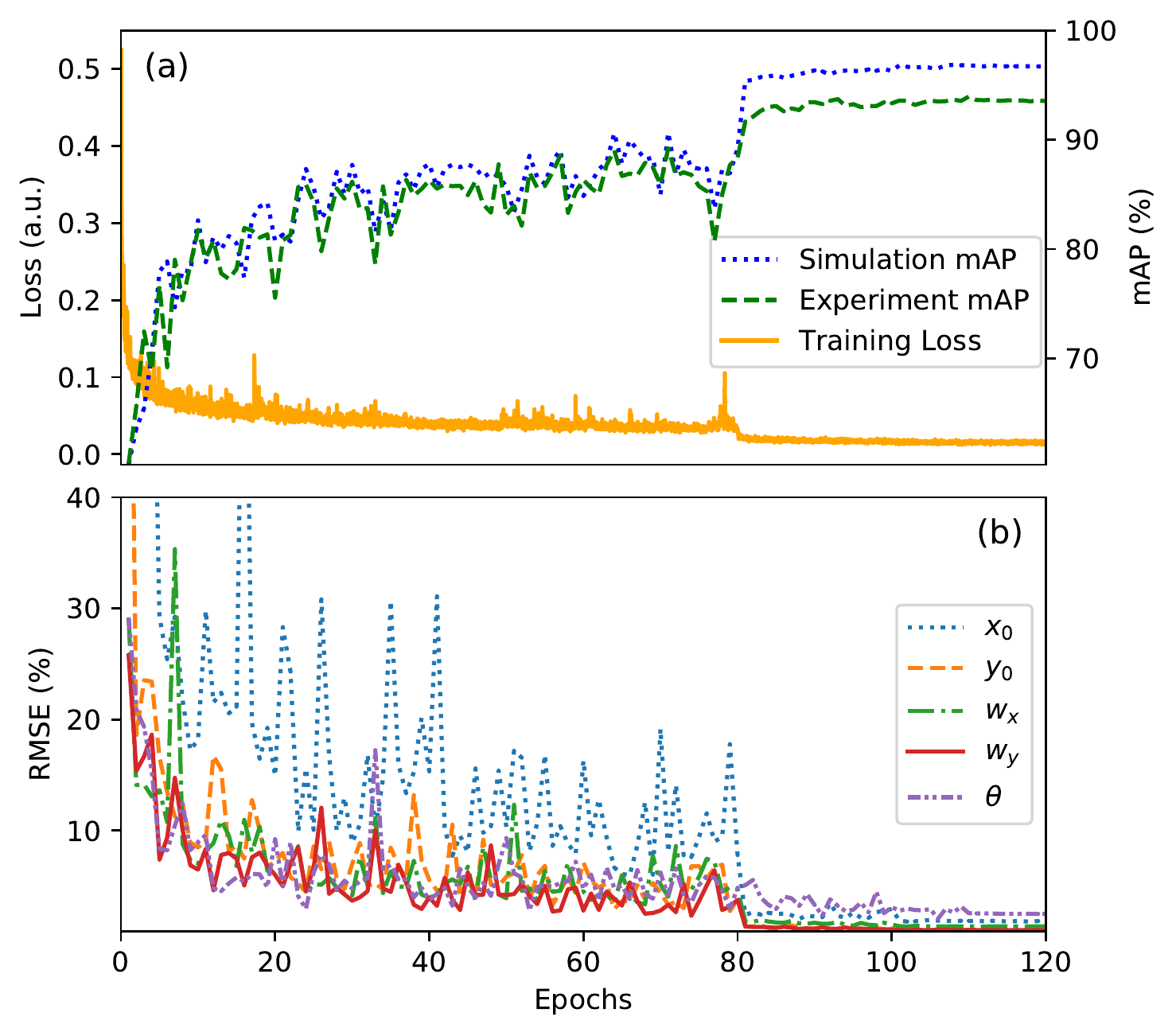}
  \caption{\label{odnn:fig:sim_data_results}Training on the simulated dataset. (a) The loss on the simulated training dataset and the mean average precision (mAP) on the simulated and experimental validation datasets plotted against the training epoch. (b) The root-mean-square errors (RMSE) for the simulation validation dataset beam parameters \{$x_0$, $y_0$, $w_x$, $w_y$, $\theta$ \} vs. the training epoch. The spatial parameters are normalized by the beam size (see text), and the angular orientation is normalized by $\pi$.}
\end{figure}

After each training epoch, the NN is evaluated on both the simulated and experimental validation datasets using the mean average precision (mAP) metric \cite{lin2014microsoft}. The mAP is a standard object detection metric where intersection-over-union scores (IoU) \cite{everingham2010pascal} between the ground truth and NN predicted RROIs are used to form precision-recall curves \cite{boyd2013area} for different IoU thresholds---which are integrated and averaged to give the mAP value \cite{lin2014microsoft}. 

At the beginning of training, the loss quickly decays while the mAPs climb, whereas towards the end of training both the loss and mAP values become asymptotic (see Fig.~\ref{odnn:fig:sim_data_results}a). Maximum mAPs of 96.8\% and 93.9\% are achieved on the simulation and experimental validation datasets respectively which correspond to the NN correctly finding 2996/3000 beams \footnote{For 35 beams, the NN predicts two RROIs, but the extra predictions are easily filtered using the IoU between predictions and the NN's prediction score.} on the simulated validation dataset and 750/750 beams on the experimental validation dataset (correct prediction threshold is set as an IoU>0.5 between the ground truth and predicted RROIs). 

Along with the mAP, the beam parameter errors are calculated after each training epoch. For both the ground truth and NN, the parameters are normalized before calculating the error. The center coordinates \{$x_0$, $y_0$\} are normalized by dividing by the ground truth beam radii along the $x$ and $y$ laboratory axes, whereas the major and minor radii \{$w_x$, $w_y$\} are divided by the ground truth major and minor radii. For the angular orientation error, $\theta$ is normalized by dividing by the range of angles, $\pi$. Beams with ellipticities below $w_x$/$w_y$=1.15 are considered radially symmetric \cite{iso2005lasers} and removed from the further angular error calculations as the angle for radially symmetric beams is arbitrary.

The spatial parameter root-mean-square-errors (RMSEs) are calculated for each validation dataset. As the training epoch increases, the RMSEs decrease (see Fig.~\ref{odnn:fig:sim_data_results}b) and for the training epochs with the best mAPs the simulated validation dataset RMSEs are all less than 2.6\% (see Table~\ref{odnn:tab:rmse_error}), whereas the experimental validation dataset RMSEs are less than 3.4\%. Note that the simulated dataset contains beams which are difficult to detect/measure such as highly elliptical beams or beams with a low signal-to-noise ratio. These help the NN learn a general definition of the Gaussian beam for inference on unseen experimental data. The high accuracy of the NN on the experimental validation dataset demonstrates the validity of this approach.

\begin{table}[t!]
\centering
{\small
\begin{tabular}{l @{\hspace{2\tabcolsep}} c @{\hspace{2\tabcolsep}} c @{\hspace{2\tabcolsep}} c @{\hspace{2\tabcolsep}} c @{\hspace{2\tabcolsep}} c @{\hspace{2\tabcolsep}} c @{\hspace{2\tabcolsep}} c @{\hspace{2\tabcolsep}} c @{\hspace{2\tabcolsep}}  c}
\hline \hline\multirow{2}{*}{Train} & \multirow{2}{*}{Val} & \multirow{2}{*}{mAP (\%)} & \multicolumn{5}{c}{RMSE (\%)} \\ \cline{4-8} &  &  & $x_0$ & $y_0$ & $w_x$ & $w_y$ & $\theta$ \\ \hline
Sim. & Sim. & 96.8 &  1.8 & 0.95 & 1.4 &  1.0 &  2.6 \\
Sim. & Exp. & 93.9 &  1.6 &  1.0 & 2.4 &  1.6 &  3.4 \\
Exp. & Exp. & 97.7 & 0.70 & 0.70 & 1.1 & 0.94 & 0.98 \\
\hline \hline
\end{tabular}
  \caption{\label{odnn:tab:rmse_error} Gaussian beam parameter root-mean-square-errors (normalized, see text) evaluated on both the simulated and experimental validation datasets (Val) at the training epoch with the highest mean average precision (mAP). The train column shows whether the neural network was trained on the simulation or experimental dataset.}}
\end{table}

\begin{table}[b!]
\centering
{\small
\begin{tabular}{l @{\hspace{0.5\tabcolsep}} c @{\hspace{0.5\tabcolsep}} c @{\hspace{0.5\tabcolsep}} c @{\hspace{0.5\tabcolsep}}  c}
\hline \hline
Parameters & Lower Bound & Upper Bound & Log Scale & Best Value \\
\hline
Learning Rate &       0.001 &        0.01 &       Yes &     0.0056 \\
Momentum      &         0.8 &       0.925 &        No &       0.85 \\
Decay Epoch   &           1 &          30 &        No &         21 \\
LR Decay      &        0.01 &         0.1 &        No &       0.01 \\
Batch Size    &           2 &           8 &        No &          5 \\
\hline \hline
\end{tabular}
  \caption{\label{odnn:exp_hyperparameters} Bounds and scaling used during Bayesian optimization of the neural network hyperparameters for training on the experimental dataset, along with the best set of hyperparameters found.}}
\end{table}

After training on the simulated dataset, the NN model weights are retained and the NN is trained on the experimental dataset. Bayesian optimization (BO) \cite{snoek2012practical, frazier2018tutorial} is used to tune the hyperparameters within sensible bounds (see Table~\ref{odnn:exp_hyperparameters})---this time with a single learning rate decay---using Facebook's Ax/BoTorch \cite{balandat2019botorch} package. Five Sobol \cite{sobol1967distribution} evaluations are used to initialize the BO loop after which a Gaussian process iteratively determines the hyperparameters for the remaining ten evaluations. For each BO evaluation, the NN is trained for thirty epochs, which still allows high accuracies to be reached due to pre-training on the simulated dataset.

Similar to the simulated training run, both the mAP and RMSEs are calculated after every training epoch on the experimental validation dataset. The NN trained with the best set of hyperparameters achieves a mAP of 97.7\% and successfully detects all 750 beams. Furthermore, the Gaussian parameter RMSEs are all below 1.1\%  (see Table~\ref{odnn:tab:rmse_error}), which are lower than the NN's experimental validation dataset RMSEs when trained on the simulated dataset alone; however, the accuracy gain is not substantial.

\section{Conclusion}

The method developed uses a deep neural network to detect an arbitrary number of Gaussian laser beams in an image and simultaneously measure their spatial parameters. The NN requires a single pass on an image which significantly simplifies the beam analysis pipeline compared to other methods. Since training the NN on simulated data alone results in high accuracies on experimental data, this method can be applied in a wide range of experimental settings. 

The NN can be used alone or can compliment other beam measurement methods by detecting laser beams in an image and determining ROIs for further calculations, such as a 2D fit or second moment measurement. This removes the need for iterative ROI algorithms which can generally only find a single beam. Furthermore, if 2D fitting is implemented, the beam parameters extracted with the NN can be used to seed the fit which increases fitting speed and the likelihood of fit convergence.

Although this method is applied to TEM$_{00}$ beams, it can be extended to higher-order Gaussian modes since RRPN natively handles multiple object types. However, this would require the training datasets to include higher-order beams with a label for each beam mode. Furthermore, for higher-order and multi-modal beams the radius is determined numerically rather than analytically---which would need to be accounted for when generating the datasets' RROI annotations.

\section*{Funding}
This work was supported by EPSRC Grant Nos. EP/P009565/1 and EP/TO19913/1, the John Fell Oxford University
Press (OUP) Research Fund and the Royal Society.

\section*{Acknowledgements}
L.H. thanks Maximilian Pfl{\"u}ger for helpful discussions.

\section*{Disclosures}

L.H. was previously employed at DataRay Inc.

\section*{Data Availability}

The data that support the findings of this study are openly available at the following URL/DOI: \url{https://doi.org/10.5287/bodleian:JbDXrnQN1}. We additionally make code available at \url{https://github.com/Dipolar-Quantum-Gases/nn-beam-profiling}.

\balance
\bibliography{./references}

\providecommand{\newblock}{}
\begin{thebibliography}{10}
\expandafter\ifx\csname url\endcsname\relax
  \def\url#1{{\tt #1}}\fi
\expandafter\ifx\csname urlprefix\endcsname\relax\def\urlprefix{URL }\fi
\providecommand{\eprint}[2][]{\url{#2}}

\bibitem{konforti1988phase}
Konforti N, Marom E and Wu S~T 1988 {\em Optics Letters\/} {\bf 13} 251--253
  \urlprefix\url{https://doi.org/10.1364/OL.13.000251}

\bibitem{Barredo}
Barredo D, de~Léséleuc S, Lienhard V, Lahaye T and Browaeys A 2016 {\em
  Science\/} {\bf 354} 1021--1023
  \urlprefix\url{https://doi.org/10.1126/science.aah3778}

\bibitem{Mello}
Ohl~de Mello D, Sch\"affner D, Werkmann J, Preuschoff T, Kohfahl L, Schlosser M
  and Birkl G 2019 {\em Physical Review Letters\/} {\bf 122}(20) 203601
  \urlprefix\url{https://doi.org/10.1103/PhysRevLett.122.203601}

\bibitem{Endres}
Endres M, Bernien H, Keesling A, Levine H, Anschuetz E~R, Krajenbrink A, Senko
  C, Vuletic V, Greiner M and Lukin M~D 2016 {\em Science\/} {\bf 354}
  1024--1027 \urlprefix\url{https://doi.org/10.1126/science.aah3752}

\bibitem{Nikolenko}
Nikolenko V, Watson B, Araya R, Woodruff A, Peterka D and Yuste R 2008 {\em
  Frontiers in Neural Circuits\/} {\bf 2} ISSN 1662-5110
  \urlprefix\url{https://doi.org/10.3389/neuro.04.005.2008}

\bibitem{ma13132962}
Hauschwitz P, Stoklasa B, Kuchařík J, Turčičová H, Písařík M, Brajer J,
  Rostohar D, Mocek T, Duda M and Lucianetti A 2020 {\em Materials\/} {\bf 13}
  ISSN 1996-1944 \urlprefix\url{https://doi.org/10.3390/ma13132962}

\bibitem{katz2018using}
Katz S, Kaplan N and Grossinger I 2018 {\em Optik \& Photonik\/} {\bf 13}
  83--86 \urlprefix\url{https://doi.org/10.1002/latj.201800021}

\bibitem{tanghetti2016histology}
Tanghetti E~A 2016 {\em Lasers in Surgery and Medicine\/} {\bf 48} 646--652
  \urlprefix\url{http://doi.org/10.1002/lsm.22540}

\bibitem{lee2019pattern}
Lee H~C, Childs J, Chung H~J, Park J, Hong J and Cho S~B 2019 {\em Scientific
  Reports\/} {\bf 9} 1--10
  \urlprefix\url{https://doi.org/10.1038/s41598-019-41021-7}

\bibitem{krizhevsky2012imagenet}
Krizhevsky A, Sutskever I and Hinton G~E 2012 {\em Advances in Neural
  Information Processing Systems\/} {\bf 25} 1097--1105
  \urlprefix\url{https://doi.org/10.1145/3065386}

\bibitem{resnet}
{He} K, {Zhang} X, {Ren} S and {Sun} J 2016 {\em IEEE Conference on Computer
  Vision and Pattern Recognition\/}  770--778
  \urlprefix\url{https://doi.org/10.1109/CVPR.2016.90}

\bibitem{doster2017machine}
Doster T and Watnik A~T 2017 {\em Applied Optics\/} {\bf 56} 3386--3396
  \urlprefix\url{https://doi.org/10.1364/AO.56.003386}

\bibitem{hofer1}
Hofer L~R, Jones L~W, Goedert J~L and Dragone R~V 2019 {\em Journal of the
  Optical Society of America A\/} {\bf 36} 936--943
  \urlprefix\url{https://doi.org/10.1364/JOSAA.36.000936}

\bibitem{lohani2018use}
Lohani S, Knutson E~M, O’Donnell M, Huver S~D and Glasser R~T 2018 {\em
  Applied Optics\/} {\bf 57} 4180--4190
  \urlprefix\url{https://doi.org/10.1364/AO.57.004180}

\bibitem{an2020fast}
An Y, Hou T, Li J, Huang L, Leng J, Yang L and Zhou P 2020 {\em Applied
  Optics\/} {\bf 59} 1954--1959
  \urlprefix\url{https://doi.org/10.1364/AO.377189}

\bibitem{Schiworski21}
Schiworski M~G, Brown D~D and Ottaway D~J 2021 {\em Journal of the Optical
  Society of America A\/} {\bf 38} 1603--1611
  \urlprefix\url{http://doi.org/10.1364/JOSAA.428214}

\bibitem{lin2018application}
Lin C~S, Huang Y~C, Chen S~H, Hsu Y~L and Lin Y~C 2018 {\em Applied Sciences\/}
  {\bf 8} 1542 \urlprefix\url{https://doi.org/10.3390/app8091542}

\bibitem{fasterrcnn}
Ren S, He K, Girshick R and Sun J 2015 {\em Advances in Neural Information
  Processing Systems\/} {\bf 28} 91--99
  \urlprefix\url{https://proceedings.neurips.cc/paper/2015/file/14bfa6bb14875e45bba028a21ed38046-Paper.pdf}

\bibitem{redmon2018yolov3}
Redmon J and Farhadi A 2018 {\em arXiv preprint arXiv:1804.02767\/}
  \urlprefix\url{https://arxiv.org/abs/1804.02767}

\bibitem{hofer2021atom}
Hofer L~R, Krstaji{\'{c}} M, Juh{\'{a}}sz P, Marchant A~L and Smith R~P 2021
  {\em Machine Learning: Science and Technology\/} {\bf 2} 045008
  \urlprefix\url{https://doi.org/10.1088/2632-2153/abf5ee}

\bibitem{rrpnn}
Ma J, Shao W, Ye H, Wang L, Wang H, Zheng Y and Xue X 2018 {\em IEEE
  Transactions on Multimedia\/} {\bf 20} 3111--3122
  \urlprefix\url{https://doi.org/10.1109/TMM.2018.2818020}

\bibitem{Gaunt2013}
Gaunt A~L, Schmidutz T~F, Gotlibovych I, Smith R~P and Hadzibabic Z 2013 {\em
  Physical Review Letters\/} {\bf 110}
  \urlprefix\url{https://doi.org/10.1103/PhysRevLett.110.200406}

\bibitem{mccally1984measurement}
McCally R~L 1984 {\em Applied Optics\/} {\bf 23} 2227--2227
  \urlprefix\url{https://doi.org/10.1364/AO.23.002227}

\bibitem{siegman1991choice}
Siegman A~E, Sasnett M and Johnston T 1991 {\em IEEE Journal of Quantum
  Electronics\/} {\bf 27} 1098--1104
  \urlprefix\url{https://doi.org/10.1109/3.83346}

\bibitem{siegman1998maybe}
Siegman A~E 1998 {\em DPSS (Diode Pumped Solid State) Lasers: Applications and
  Issues\/}  MQ1 \urlprefix\url{https://doi.org/10.1364/DLAI.1998.MQ1}

\bibitem{ross2013laser}
Ross T~S 2013 {\em Laser beam quality metrics\/} (SPIE Press)
  \urlprefix\url{https://doi.org/10.1117/3.1000595}

\bibitem{iso2005lasers}
 2005 Iso 11146-1: Lasers and laser-related equipment—test methods for laser
  beam widths, divergence angles and beam propagation ratios—part 1:
  Stigmatic and simple astigmatic beams

\bibitem{Hofer2017}
Hofer L~R, Dragone R~V and MacGregor A~D 2017 {\em Optical Engineering\/} {\bf
  56} 043110 \urlprefix\url{https://doi.org/10.1117/1.OE.56.4.043110}

\bibitem{girshick2014rich}
{Girshick} R, {Donahue} J, {Darrell} T and {Malik} J 2014 {\em {IEEE}
  {Conference} on {Computer} {Vision} and {Pattern} {Recognition}\/}  580--587
  \urlprefix\url{https://doi.org/10.1109/CVPR.2014.81}

\bibitem{zeiler2014visualizing}
Zeiler M~D and Fergus R 2014 {\em Computer Vision -- ECCV\/}  818--833
  \urlprefix\url{https://doi.org/10.1007/978-3-319-10590-1}

\bibitem{huang2018improving}
Huang J, Sivakumar V, Mnatsakanyan M and Pang G 2018 {\em arXiv preprint
  arXiv:1811.07031\/} \urlprefix\url{https://arxiv.org/abs/1811.07031}

\bibitem{wood2016learning}
Wood E, Baltru{\v{s}}aitis T, Morency L~P, Robinson P and Bulling A 2016 {\em
  Proceedings of the Ninth Biennial ACM Symposium on Eye Tracking Research \&
  Applications\/}  131--138
  \urlprefix\url{https://doi.org/10.1145/2857491.2857492}

\bibitem{yoo2016pixel}
Yoo D, Kim N, Park S, Paek A~S and Kweon I~S 2016 {\em European Conference on
  Computer Vision\/}  517--532
  \urlprefix\url{http://doi.org/10.1007/978-3-319-46484-8_31}

\bibitem{zhang2015learning}
Zhang X, Fu Y, Zang A, Sigal L and Agam G 2015 {\em arXiv preprint
  arXiv:1503.03163\/} \urlprefix\url{https://arxiv.org/abs/1503.03163}

\bibitem{forbes2019structured}
Forbes A 2019 {\em Laser \& Photonics Reviews\/} {\bf 13} 1900140
  \urlprefix\url{https://doi.org/10.1002/lpor.201970043}

\bibitem{Forbes16}
Forbes A, Dudley A and McLaren M 2016 {\em Advances in Optics and Photonics\/}
  {\bf 8} 200--227 \urlprefix\url{https://doi.org/10.1364/AOP.8.000200}

\bibitem{arrizon2007pixelated}
Arriz{\'o}n V, Ruiz U, Carrada R and Gonz{\'a}lez L~A 2007 {\em Journal of the
  Optical Society of America A\/} {\bf 24} 3500--3507
  \urlprefix\url{https://doi.org/10.1364/JOSAA.24.003500}

\bibitem{rosales2017shape}
Rosales-Guzm{\'a}n C and Forbes A 2017 {\em How to shape light with spatial
  light modulators\/} (SPIE Press)
  \urlprefix\url{https://doi.org/10.1117/3.2281295}

\bibitem{wu2019detectron2}
Wu Y, Kirillov A, Massa F, Lo W~Y and Girshick R 2019 Detectron2
  \url{https://github.com/facebookresearch/detectron2}

\bibitem{yosinski2014transferable}
Yosinski J, Clune J, Bengio Y and Lipson H 2014 {\em Advances in Neural
  Information Processing Systems\/} {\bf 27} 3320--3328
  \urlprefix\url{https://proceedings.neurips.cc/paper/2014/file/375c71349b295fbe2dcdca9206f20a06-Paper.pdf}

\bibitem{bergstra2012random}
Bergstra J and Bengio Y 2012 {\em The Journal of Machine Learning Research\/}
  {\bf 13} 281--305
  \urlprefix\url{https://dl.acm.org/doi/abs/10.5555/2188385.2188395}

\bibitem{bisong2019google}
Bisong E 2019 {\em Google {Colaboratory}\/} (Springer) pp 59--64
  \urlprefix\url{https://link.springer.com/book/10.1007/978-1-4842-4470-8}

\bibitem{lin2014microsoft}
Lin T~Y, Maire M, Belongie S, Hays J, Perona P, Ramanan D, Doll{\'a}r P and
  Zitnick C~L 2014 {\em Computer Vision -- ECCV\/}  740--755
  \urlprefix\url{https://doi.org/10.1007/978-3-319-10602-1}

\bibitem{everingham2010pascal}
Everingham M, Van~Gool L, Williams C~K, Winn J and Zisserman A 2010 {\em
  International Journal of Computer Vision\/} {\bf 88} 303--338
  \urlprefix\url{https://doi.org/10.1007/s11263-009-0275-4}

\bibitem{boyd2013area}
Boyd K, Eng K~H and Page C~D 2013 {\em Joint European Conference on Machine
  Learning and Knowledge Discovery in Databases\/}  451--466
  \urlprefix\url{http://doi.org/10.1007/978-3-642-40994-3}

\bibitem{snoek2012practical}
Snoek J, Larochelle H and Adams R~P 2012 {\em Advances in Neural Information
  Processing Systems\/} {\bf 25} 2951--2959
  \urlprefix\url{https://proceedings.neurips.cc/paper/2012/file/05311655a15b75fab86956663e1819cd-Paper.pdf}

\bibitem{frazier2018tutorial}
Frazier P~I 2018 {\em arXiv preprint arXiv:1807.02811\/}
  \urlprefix\url{https://arxiv.org/abs/1807.02811}

\bibitem{balandat2019botorch}
Balandat M, Karrer B, Jiang D~R, Daulton S, Letham B, Wilson A~G and Bakshy E
  2020 {\em arXiv preprint arXiv:1910.06403\/}
  \urlprefix\url{https://arxiv.org/abs/1910.06403}

\bibitem{sobol1967distribution}
Sobol' I~M 1967 {\em Zhurnal Vychislitel'noi Matematiki i Matematicheskoi
  Fiziki\/} {\bf 7} 784--802 \urlprefix\url{http://mi.mathnet.ru/eng/zvmmf7334}

\end{thebibliography}

\end{document}